\documentclass[a4paper]{article}
\usepackage{amsfonts}
\usepackage{amsmath}

\begin{document}

\title{General $SU(2)_L\times SU(2)_R$ $\times U(1)_{EM}$ $\ $Sigma Model With
External Sources, Dynamical Breaking And Spontaneous Vacuum Symmetry
Breaking }
\author{Yong-Chang Huang$^{1,3}$ Xi-Guo Lee$^{2,4}$ Liu-Ji Li$^1$  \\
%EndAName
$^1$Institute of Theoretical Physics, Beijing University of Technology,
Beijing 100022, P. R. China\\
$^2$Institute of Modern Physics, Chinese Academy of Science, Lanzhou,
730000, P. R. China\\
$^3$CCAST ( World Lab. ), P. O. Box 8730, Beijing, 100080, P. R. China\\
$^4$Center of Theoretical Nuclear Physics, National Laboratory of Heavy Ion
Collisions,\\
Lanzhou 730000, P. R. China\\
PACA Numbers: 24.10.-i, 11.30.Qc}
\maketitle

\begin{abstract}
We give a general $SU(2)_L\times SU(2)_R$ $\times U(1)_{EM}$ sigma model
with external sources, dynamical breaking and spontaneous vacuum symmetry
breaking, and present the general formulation of the model. It is found that
$\sigma $ and $\pi ^0$ without electric charges have electromagnetic
interaction effects coming from their internal structure. A general Lorentz
transformation relative to external sources $J_{gauge}$ $=(J_{A_\mu
},J_{A_\mu ^\kappa })$ is derived, using the general Lorentz transformation
and the four-dimensional current of nuclear matter of the ground state with $%
J_{gauge}$ = 0, we give the four-dimensional general relations between the
different currents of nuclear matter systems with $J_{gauge}\neq 0$ and
those with $J_{gauge}=0$. The relation of the density's coupling with
external magnetic field is derived , which conforms well to dense nuclear
matter in a strong magnetic field. We show different condensed effects in
strong interaction about fermions and antifermions, and give the concrete
scalar and pseudoscalar condensed expressions of $\sigma _0$ and $\pi _0$
bosons. About different dynamical breaking and spontaneous vacuum symmetry
breaking, the concrete expressions of different mass spectra are obtained in
field theory. This paper acquires the running spontaneous vacuum breaking
value $\sigma _0^{\prime },$ and obtains the spontaneous vacuum breaking in
terms of the running $\sigma _0^{\prime }$, which make nucleon, $\sigma $
and $\pi $ particles gain effective masses. We achieve both the effect of
external sources and nonvanishing value of the condensed scalar and
pseudoscalar paticles. It is deduced that the masses of nucleons, $\sigma $
and $\pi $ generally depend on different external sources.
\end{abstract}

\section{Introduction}

It is well known that a lot of unified models of the electroweak and the
strong interactions utilize symmetry breaking. Now it is widely believed
that the underlying laws of the world have essential symmetries\cite{wei}.
It is that symmetric equations may have asymmetric solutions. And also there
are various symmetry breaking in the world\cite{and,nam}. Dynamical symmetry
breaking of extended gauge symmetries is given\cite{thom}. The Ref.\cite
{thoms} shows dynamical symmetry breaking in the sea of the nucleon, and the
Ref.\cite{stet} investigates dynamical electroweak symmetry breaking by a
neutrino condensate.

Spontaneous symmetry breaking plays an important role for constructing the
different unified theories of the electroweak and the strong interactions,
as well as gravity theory\cite{mig}. But the fundamental scalar field, e.g.
Higgs particle, has not been discovered up to now, even though the low
energy limit of finding Higgs particle has been raised to very high\cite{acc}%
, especially in testing the standard model of the weak-electromagnetic
interactions. The different grand unified theories have many parameters
adjusted to fit the experiments, which make the theoretical predication to
physical properties be decreased. On the other hand, there are other
mechanisms generating particle's masses [9-12]. The Ref.\cite{sch} indicates
that if the vacuum polarization tensor has a pole at light-like momenta,
gauge field may acquire mass. A classical $\sigma $ model of chiral symmetry
breaking was given in Ref. \cite{namb}, and an in-medium QMC model
parameterization quark condensation in nuclear matter etc are studied in Ref%
\cite{guo,guoh}.

The pure interactions mediated by swapped mesons between fermions and
antifermions possibly yield vacuum condensation of the fermion-antifermions
pair \cite{liu}, which makes vacuum degeneracy appears. Ref. \cite{civi}
researched spontaneous and dynamical breaking of mean field symmetries in
the proton-neutron quasiparticle random phase approximation and the
description of double beta decay transitions. And dynamical chiral symmetry
breaking in gauge theories with extra dimensions is also well described\cite
{gusy}.

Dynamical electroweak breaking and latticized extra dimensions are shown up%
\cite{hsi}, using dynamical breaking, one may make fermions and bosons get
masses, and may make the free adjusted parameters decrease, even to a
dynamical group of one parameter.

When considering the physical effect of a system coming from another system,
a general quantitative causal conservation principle must be satisfied\cite
{hpri}. Using the homeomorphic map transformation satisfying the general
quantitative causal conservation principle, Ref.\cite{hcon} solves the hard
problem of the non-perfect properties of the Volterra process, the
topological current invariants in Riemann-Cartan manifold and spacetime
defects still satisfy the general quantitative causal conservation principle%
\cite{huan}. This paper illustrates the fact that $\sigma $ and $\pi ^0$
without electric charges having electromagnetic interaction effects coming
from their inner constructions are just result satisfying the general causal
conservation rule, i.e., the general quantitative causal conservation
principle is essential for researching consistency of the model.

In general analyzing vacuum degeneracy, one studies only the degeneracy
vacuum state originated from the self-action of scalar fields, one usually
neglects the vacuum degeneracy originated from the interactions of different
fields.

In this paper, Sect.2 gives the basic formulation; Sect.3 studies different
condensation about fermions and antifermions; Sect.4 gives the concrete
expressions of different mass spectrum about different vacuum breaking and
dynamical breaking, and shows that the general four dimensional relations
between different currents of the nuclear matter systems with J$\neq 0$ and
those with $J=0$; the last Sect. is summary and conclusion.

\section{Basic Formulation}

The Lagrangian of the general $\sigma $-model with the symmetries of chiral $%
SU(2)_L\times SU(2)_R$ and electromagnetic $U(1)_{EM}$ is
\begin{equation}
\frak{L}_j=\frak{L}+\overline{\eta }\psi +\overline{\psi }\eta +J_\sigma
\sigma +\mathbf{J}_\pi \cdot \mathbf{\pi }+J_{A_\mu }A_\mu .  \tag{2.2}
\end{equation}

Euler-Lagrange Equations of the system are

\begin{equation}
\lbrack \gamma ^\mu \partial _\mu -ieA_\mu +g(\sigma (x)+i\mathbf{\tau }%
\cdot \mathbf{\pi }(x)\gamma _5)]\psi (x)-\eta (x)=0,  \tag{2.3}
\end{equation}

\begin{equation}
\overline{\psi }(x)[-\gamma ^\mu \stackrel{\leftarrow }{\partial _\mu }%
-ie\gamma ^\mu A_\mu +g(\sigma (x)+i\mathbf{\tau }\cdot \mathbf{\pi }%
(x)\gamma _5)]-\overline{\eta }(x)=0 ,  \tag{2.4}
\end{equation}

\begin{equation}
(\Box +\lambda \nu ^2)\sigma (x)-g\overline{\psi }(x)\psi (x)-\lambda \sigma
(x)(\sigma ^2(x)+\mathbf{\pi }^2(x))+J_\sigma (x)=0,  \tag{2.5}
\end{equation}

\begin{equation}
(\Box +\lambda \nu ^2)\mathbf{\pi }(x)-e^2A_\mu ^2(x)\mathbf{\pi }(x)-g%
\overline{\psi }(x)i\mathbf{\tau }\gamma _5\psi (x)-\lambda \mathbf{\pi }%
(x)(\sigma ^2(x)+\mathbf{\pi }^2(x))+\mathbf{J}_\pi (x)=0 ,  \tag{2.6}
\end{equation}
and
\begin{equation}
\partial _\nu F^{\mu \nu }+ie\overline{\psi }(x)\gamma ^\mu \psi
(x)-e^2A_\mu (x)\mathbf{\pi }^2(x)+J_{A_\mu }=0.  \tag{2.7}
\end{equation}

Then we have

\begin{equation}
\langle \overline{\psi }(x)\gamma ^\mu \partial _\mu \psi (x)\rangle
_0^J-ie\langle \overline{\psi }(x)A_\mu (x)\psi (x)\rangle _0^J+g\langle
\overline{\psi }(x)\sigma (x)\psi (x)\rangle _0^J+i\langle \overline{\psi }%
(x)\mathbf{\tau }\cdot \mathbf{\pi }(x)\gamma _5\psi (x)\rangle _0^J-\langle
\overline{\psi }(x)\rangle _0^J\eta (x)=0  \tag{2.8}
\end{equation}
and
\begin{equation}
\langle \overline{\psi }(x)\gamma ^\mu \stackrel{\leftarrow }{\partial \mu }%
\psi (x)\rangle _0^J+ie\langle \overline{\psi }(x)A_\mu (x)\psi (x)\rangle
_0^J-g\langle \overline{\psi }(x)\sigma (x)\psi (x)\rangle _0^J-i\langle
\overline{\psi }(x)\mathbf{\tau }\cdot \mathbf{\pi }(x)\gamma _5\psi
(x)\rangle _0^J+\overline{\eta }(x)\langle \psi (x)\rangle _0^J=0.  \tag{2.9}
\end{equation}

We can further obtain

\begin{equation}
(\Box +\lambda \nu ^2)\langle \sigma (x)\rangle _0^J-g\langle \overline{\psi
}(x)\psi (x)\rangle _0^J-\lambda \langle \sigma (x)(\sigma ^2(x)+\mathbf{\pi
}^2(x)\rangle _0^J+J_\sigma (x)=0 ,  \tag{2.10}
\end{equation}

\begin{equation}
(\Box +\lambda \nu ^2)\langle \mathbf{\pi }(x)\rangle _0^J-e^2\langle A_\mu
^2(x)\mathbf{\pi }(x)\rangle _0^J-g\langle \overline{\psi }(x)i\mathbf{\tau }%
\gamma _5\psi (x)\rangle _0^J-\lambda \langle \mathbf{\pi }(x)(\sigma ^2(x)+%
\mathbf{\pi }^2(x))\rangle _0^J+\mathbf{J}_\pi (x)=0,  \tag{2.11}
\end{equation}

\begin{equation}
\left\langle \partial _\nu F^{\mu \nu }\right\rangle _0^J+ie\langle
\overline{\psi }(x)\gamma ^\mu \psi (x)\rangle _0^J-e^2\langle A_\mu (x)%
\mathbf{\pi }^2(x)\rangle _0^J+J_{A_\mu }(x)=0,  \tag{2.12}
\end{equation}
in which for any field, we can define $\langle Y(x)\rangle _0^J\equiv
\langle 0_{out}\left| Y(x)\right| 0_{in}\rangle _0^J\left/ \langle
0_{out}\right| 0_{in}\rangle _0^J$.

The generating functional of the system is

\begin{equation}
Z(J)\equiv \int \left[ D\overline{\psi }\right] \left[ D\psi \right] \left[
D\sigma \right] \left[ D\mathbf{\pi }\right] \left[ DA_\mu \right] \text{exp}%
\left( \frac i\hbar \int d^4x\frak{L}_J\right)  \tag{2.13}
\end{equation}
Using the generating function one have

\begin{equation}
\langle Y(x)\rangle _0^J=\hbar \frac{\delta W}{\delta J_Y(x)},  \tag{2.14}
\end{equation}
where $Z=e^{iW}$.

On the other hand, using the method of deducing connection Green function
from Green function in quantum field theory\cite{cla,ygau} we can have

\begin{equation}
\langle \sigma ^3(x)\rangle _0^J=(\langle \sigma (x)\rangle _0^J)^3+3\frac
\hbar i\langle \sigma (x)\rangle _0^J\frac{\delta \langle \sigma (x)\rangle
_0^J}{\delta J_\sigma (x)}+(\frac \hbar i)^2\frac{\delta ^2\langle \sigma
(x)\rangle _0^J}{\delta J_\sigma (x)\delta J_\sigma (x)}+\cdot \cdot \cdot ,
\tag{2.15}
\end{equation}

\begin{align}
\langle \sigma (x)\mathbf{\pi }^2(x)\rangle _0^J &=(\langle \mathbf{\pi }%
^2(x)\rangle _0^J)^2\langle \sigma (x)\rangle _0^J+\frac \hbar i\langle
\sigma (x)\rangle _0^J\frac{\delta \langle \mathbf{\pi }(x)\rangle _0^J}{%
\delta \mathbf{J}_\pi (x)}  \tag{2.16}\\
&+2\frac \hbar i\langle \mathbf{\pi }(x)\rangle _0^J\cdot
\frac{\delta \langle \Bbb{\sigma }(x)\rangle _0^J}{\delta
\mathbf{J}_\pi (x)}+(\frac \hbar i)^2\frac{\delta ^2\langle
\Bbb{\sigma }(x)\rangle _0^J}{\delta \mathbf{J}_\pi (x)\cdot \delta
\mathbf{J}_\pi (x)}+\cdot \cdot \cdot , \nonumber
\end{align}

\begin{align}
\langle \mathbf{\pi }(x)A_\mu ^2(x)\rangle _0^J &=&(\langle A_\mu (x)\rangle
_0^J)^2\langle \mathbf{\pi }(x)\rangle _0^J+\frac \hbar i\langle \mathbf{\pi
}(x)\rangle _0^J\frac{\delta \langle A_\mu (x)\rangle _0^J}{\delta J_{A_\mu
}(x)}  \tag{2.17} \\
&&+2\frac \hbar i\langle A_\mu (x)\rangle _0^J\frac{\delta \langle \mathbf{%
\pi }(x)\rangle _0^J}{\delta J_{A_\mu }(x)}+(\frac \hbar i)^2\frac{\delta
^2\langle \mathbf{\pi }(x)\rangle _0^J}{\delta J_{A_\mu }(x)\cdot \delta
J_{A_\mu }(x)}+\cdot \cdot \cdot ,  \nonumber
\end{align}

\begin{align}
\langle A_\mu (x)\mathbf{\pi }^2(x)\rangle _0^J &=&\langle A_\mu (x)\rangle
_0^J(\langle \mathbf{\pi }(x)\rangle _0^J)^2+\frac \hbar i\langle A_\mu
(x)\rangle _0^J\frac{\delta \langle \mathbf{\pi }(x)\rangle _0^J}{\delta
\mathbf{J}_\pi (x)}  \tag{2.18} \\
&&+2\frac \hbar i\langle \mathbf{\pi }(x)\rangle _0^J\cdot \frac{\delta
\langle A_\mu (x)\rangle _0^J}{\delta \mathbf{J}_\pi (x)}+(\frac \hbar i)^2%
\frac{\delta ^2\langle A_\mu (x)\rangle _0^J}{\delta \mathbf{J}_\pi (x)\cdot
\delta \mathbf{J}_\pi (x)}+\cdot \cdot \cdot ,  \nonumber
\end{align}
which are just a kind of new power expansion about the little quantity $%
\hbar $, which is essential for researching the physics of different power
series about $\hbar .$ Because there are possible condensations of $\langle
\overline{\psi }(x)\psi (x)\rangle _0^J$, $\langle \overline{\psi }(x)i%
\mathbf{\tau }\Bbb{\gamma }_5\psi (x)\rangle _{0\text{ }}^J$and $\langle
\overline{\psi }(x)\Bbb{\gamma }_\mu \psi (x)\rangle _{0\text{ }}^J$in Eqs.
(2.10-12), respectively, we have

\begin{equation}
\langle \overline{\psi }(x)\sigma (x)\psi (x)\rangle _0^J=\langle \sigma
(x)\rangle _0^J\langle \overline{\psi }(x)\psi (x)\rangle _0^J+\frac \hbar i%
\frac{\delta \langle \overline{\psi }(x)\psi (x)\rangle _0^J}{\delta
J_\sigma (x)}+\cdot \cdot \cdot,  \tag{2.19}
\end{equation}

\begin{equation}
\langle \overline{\psi }(x)i\mathbf{\tau \cdot \pi }(x)\Bbb{\gamma }_5\psi
(x)\rangle _0^J=\langle \mathbf{\pi }(x)\rangle _0^J\cdot \langle \overline{%
\psi }(x)i\mathbf{\tau }\Bbb{\gamma }_5\psi (x)\rangle _0^J+\frac \hbar i%
\frac{\delta \langle \overline{\psi }(x)i\mathbf{\tau }\Bbb{\gamma }_5\psi
(x)\rangle _0^J}{\delta \mathbf{J}_\pi (x)}+\cdot \cdot \cdot ,  \tag{2.20}
\end{equation}

\begin{equation}
\langle \overline{\psi }(x)A_\mu (x)\Bbb{\gamma }_\mu \psi (x)\rangle
_0^J=\langle A_\mu (x)\rangle _0^J\langle \overline{\psi }(x)\Bbb{\gamma }%
_\mu \psi (x)\rangle _0^J+\frac \hbar i\frac{\delta \langle \overline{\psi }%
(x)\Bbb{\gamma }_\mu \psi (x)\rangle _0^J}{\delta J_{A_\mu }(x)}+\cdot \cdot
\cdot \text{ }.  \tag{2.21}
\end{equation}
Hence, we obtain

\[
\langle \overline{\psi }(x)\gamma ^\mu \partial \mu \psi (x)\rangle
_0^J-ie\langle A_\mu (x)\rangle _0^J\langle \overline{\psi }(x)\gamma ^\mu
\psi (x)\rangle _0^J+g\langle \sigma (x)\rangle _0^J\langle \overline{\psi }%
(x)\psi (x)\rangle _0^J+ig\langle \pi (x)\rangle _0^J\langle \overline{\psi }%
(x)\mathbf{\tau }\Bbb{\gamma }_5\psi (x)\rangle _0^J
\]

\begin{equation}
-\langle \overline{\psi }(x)\rangle _0^J\eta (x)-e\hbar \frac{\delta \langle
\overline{\psi }(x)\psi (x)\rangle _0^J}{\delta J_{A_\mu }(x)}+g\frac \hbar i%
\frac{\delta \langle \overline{\psi }(x)\psi (x)\rangle _0^J}{\delta
J_\sigma (x)}+g\hbar \frac{\delta \langle \overline{\psi }(x)\psi (x)\rangle
_0^J}{\delta \mathbf{J}_\pi (x)}+\cdot \cdot \cdot =0,  \tag{2.22}
\end{equation}

\[
-\langle \overline{\psi }(x)\gamma ^\mu \stackrel{\leftarrow }{\partial \mu }%
\psi (x)\rangle _0^J-ie\langle A_\mu (x)\rangle _0^J\langle \overline{\psi }%
(x)\gamma ^\mu \psi (x)\rangle _0^J+g\langle \sigma (x)\rangle _0^J\langle
\overline{\psi }(x)\psi (x)\rangle _0^J+ig\langle \pi (x)\rangle _0^J\langle
\overline{\psi }(x)\mathbf{\tau }\Bbb{\gamma }_5\psi (x)\rangle _0^J
\]

\begin{equation}
-\stackrel{\_}{\eta }(x)\langle \psi (x)\rangle _0^J-e\hbar \frac{\delta
\langle \overline{\psi }(x)\psi (x)\rangle _0^J}{\delta J_{A_\mu }(x)}%
+g\frac \hbar i\frac{\delta \langle \overline{\psi }(x)\psi (x)\rangle _0^J}{%
\delta J_\sigma (x)}+g\hbar \frac{\delta \langle \overline{\psi }(x)\psi
(x)\rangle _0^J}{\delta \mathbf{J}_\pi (x)}+\cdot \cdot \cdot =0,  \tag{2.23}
\end{equation}
and we can have
\[
(\Box +\lambda \nu ^2)\langle \sigma (x)\rangle _0^J=g\langle \overline{\psi
}(x)\psi (x)\rangle _0^J+\lambda \langle \sigma (x)\rangle _0^J[(\langle
\sigma (x)\rangle _0^J)^2+(\langle \mathbf{\pi }(x)\rangle _0^J)^2]+\lambda
\frac \hbar i[3\langle \sigma (x)\rangle _0^J\frac{\delta \langle \sigma
(x)\rangle _0^J}{\delta J_\sigma (x)}+
\]

\begin{equation}
\langle \sigma (x)\rangle _0^J\frac{\delta \langle \mathbf{\pi }(x)\rangle
_0^J}{\delta \mathbf{J}_\pi (x)}+2\langle \mathbf{\pi }(x)\rangle _0^J\cdot
\frac{\delta \langle \sigma (x)\rangle _0^J}{\delta \mathbf{J}_\pi (x)}%
]+\lambda (\frac \hbar i)^2[\frac{\delta ^2\langle \sigma (x)\rangle _0^J}{%
\delta J_\sigma (x)\delta J_\sigma (x)}+\frac{\delta ^2\langle \sigma
(x)\rangle _0^J}{\delta \mathbf{J}_\pi (x)\cdot \delta \mathbf{J}_\pi (x)}%
]-J_\sigma (x)+\cdot \cdot \cdot ,  \tag{2.24}
\end{equation}

\[
(\Box +\lambda \nu ^2)\langle \mathbf{\pi }(x)\rangle _0^J=g\langle
\overline{\psi }(x)i\mathbf{\tau }\Bbb{\gamma }_5\psi (x)\rangle
_0^J+\lambda \langle \mathbf{\pi }(x)\rangle _0^J[(\langle \sigma (x)\rangle
_0^J)^2+(\langle \mathbf{\pi }(x)\rangle _0^J)^2]+\lambda \frac \hbar
i[3\langle \mathbf{\pi }(x)\rangle _0^J\frac{\delta \langle \mathbf{\pi }%
(x)\rangle _0^J}{\delta \mathbf{J}_\pi (x)}
\]

\[
+2\langle \sigma (x)\rangle _0^J\frac{\delta \langle \mathbf{\pi }(x)\rangle
_0^J}{\delta J_\sigma (x)}+\langle \mathbf{\pi }(x)\rangle _0^J\cdot \frac{%
\delta \langle \sigma (x)\rangle _0^J}{\delta J_\sigma (x)}]+\lambda (\frac
\hbar i)^2[\frac{\delta ^2\langle \mathbf{\pi }(x)\rangle _0^J}{\delta
J_\sigma (x)\cdot \delta J_\sigma (x)}+\frac{\delta ^2\langle \mathbf{\pi }%
(x)\rangle _0^J}{\delta \mathbf{J}_\pi (x)\cdot \delta \mathbf{J}_\pi (x)}]-%
\mathbf{J}_\pi (x)+
\]

\begin{equation}
e^2[(\langle A_\mu (x)\rangle _0^J)^2\langle \mathbf{\pi }(x)\rangle
_0^J+\frac \hbar i\langle \mathbf{\pi }(x)\rangle _0^J\frac{\delta \langle
A_\mu (x)\rangle _0^J}{\delta J_{A_\mu }(x)}+2\frac \hbar i\langle A_\mu
(x)\rangle _0^J\frac{\delta \langle \mathbf{\pi }(x)\rangle _0^J}{\delta
J_{A_\mu }(x)}+(\frac \hbar i)^2\frac{\delta ^2\langle \mathbf{\pi }%
(x)\rangle _0^J}{\delta J_{A_\mu }(x)\delta J_{A_\mu }(x)}]+\cdots,
\tag{2.25}
\end{equation}

\begin{align}
&&\langle \partial _\nu F^{\mu \nu }\rangle _0^J+ie\langle \overline{\psi }%
(x)\Bbb{\gamma }^\mu \psi (x)\rangle _0^J-e^2[(\langle \mathbf{\pi }%
(x)\rangle _0^J)^2\langle A_\mu (x)\rangle _0^J+\frac \hbar i\langle A_\mu
(x)\rangle _0^J\frac{\delta \langle \mathbf{\pi }(x)\rangle _0^J}{\delta
\mathbf{J}_\pi (x)}  \tag{2.26} \\
&&+2\frac \hbar i\langle \mathbf{\pi }(x)\rangle _0^J\cdot \frac{\delta
\langle A_\mu (x)\rangle _0^J}{\delta \mathbf{J}_\pi (x)}+(\frac \hbar i)^2%
\frac{\delta ^2\langle \mathbf{\pi }(x)\rangle _0^J}{\delta \mathbf{J}_\pi
(x)\cdot \delta \mathbf{J}_\pi (x)}+\cdot \cdot \cdot ]+J_{A_\mu }(x).
\nonumber
\end{align}

And we can further obtain

\begin{equation}
\langle \partial _\mu (\overline{\psi }(x)\Bbb{\gamma }^\mu \psi (x))\rangle
_0^J=\langle \overline{\psi }(x)\rangle _0^J\eta (x)-\overline{\eta }%
(x)\langle \psi (x)\rangle _0^J,  \tag{2.27}
\end{equation}
when $\overline{\eta }=\eta =0,$ it follows that

\begin{equation}
\partial _\mu \overline{(\psi }(x)\Bbb{\gamma }^\mu \psi (x))=0\text{, }i.e.,%
\text{ }\partial _\mu j^\mu =0.  \tag{2.28}
\end{equation}

We neglect the powers with $\hbar $ in the power series, and take external
sources into zero, therefore, we deduce

\begin{equation}
g\langle \overline{\psi }(x)\psi (x)\rangle _0^J\mid _{J=0}+\lambda \sigma
_0(\sigma _0^2+\mathbf{\pi }_0^2-\nu ^2)=0,  \tag{2.29}
\end{equation}

\begin{equation}
ig\langle \overline{\psi }(x)\Bbb{\gamma }_5\mathbf{\tau }\psi (x)\rangle
_0^J\mid _{J=0}+\lambda \mathbf{\pi }_0(\sigma _0^2+\mathbf{\pi }_0^2-\nu
^2)=0 ,  \tag{2.30}
\end{equation}

\begin{equation}
\langle \partial _\nu F^{\mu \nu }\rangle _0^J\mid _{J=0}+ie\langle (%
\overline{\psi }(x)\Bbb{\gamma }^\mu \psi (x)\rangle _0^J\mid _{J=0}=0,
\tag{2.31}
\end{equation}
where $\sigma _0=\langle \sigma (x)\rangle _0^J\mid _{J=\text{ }0}$and $%
\mathbf{\pi }_0=\langle \mathbf{\pi (}x\mathbf{)}\rangle _0^J\mid _{J=\text{
}0}$.

Analogous to Ref.\cite{lul}$^{\prime }$s research, fermion$^{\prime }$s
propagator is

\begin{equation}
\langle \overline{\psi }(x)\psi (x^{\prime })\rangle _0^J=\frac 1{(2\pi
)^4}\int^\Lambda \frac{-e^{i(x-x^{\prime })\text{ }\cdot \ p}d^4p}{\Bbb{%
\gamma }^\mu \cdot p_\mu -ig\langle \sigma (x)\rangle _0^J+g\mathbf{\tau
\cdot }\langle \mathbf{\pi (}x\mathbf{)}\rangle _0^J\Bbb{\gamma }_5-e\Bbb{%
\gamma }^\mu \langle A_\mu (x)\rangle _0^J},  \tag{2.32}
\end{equation}
where $\Lambda $ is the cutting parameter, Eqs.(2.28-32) are the basic
equations relative to both dynamical breaking and vacuum breaking.

\section{Different Condensations About Fermions and Antifermions and the
Four Dimensional General Different Currents}

We now generally investigate the different condensations about fermions and
antifermions.

When $\sigma _0\neq 0,$ $\langle \overline{\psi }(x)\psi (x)\rangle _0^J\mid
_{J=0}\neq 0$, we evidently have

\begin{equation}
\frac{ig\langle \overline{\psi }(x)\Bbb{\gamma }_5\mathbf{\tau }\psi
(x)\rangle _0^J\mid _{J=0}}{g\langle \overline{\psi }(x)\psi (x)\rangle
_0^J\mid _{J=0}}=\frac{\mathbf{\pi }_0}{\sigma _0},  \tag{3.1}
\end{equation}
then we generally have

\begin{equation}
\sigma _0=Kg\langle \overline{\psi }(x)\psi (x)\rangle _0^J\mid _{J=0},
\tag{3.2}
\end{equation}

\begin{equation}
\mathbf{\pi }_0=iKg\langle \overline{\psi }(x)\Bbb{\gamma }_5\mathbf{\tau }%
\psi (x)\rangle _0^J\mid _{J=0},  \tag{3.3}
\end{equation}
where K is the parameter determined by physical experiments or theoretical
model. Eq.(3.2) and (3.3) mean that $\sigma _0$ and $\mathbf{\pi }_0$ are
directly originated from the dynamical condensations of fermion-antifermion.
The condensations also depend on $K,$ which is different from the
condensation mechanism before.

Analogous to Ref.\cite{dav}, it shows that under some conditions the
fundamental scalar fields are equivalent to the composed scalar fields.

Furthermore, we have

\begin{equation}
ic\langle \rho _e(x)\rangle _0^J\mid _{J=0}=\langle \partial _\nu F^{4\nu
}(x)\rangle _0^J\mid _{J=0}=-ie\langle \psi ^{+}(x)\psi (x)\rangle _0^J\mid
_{J=0},  \tag{3.4}
\end{equation}

\begin{equation}
\langle j_e^i\rangle _0^J\mid _{J=0}=\langle \partial _\nu F^{i\nu
}(x)\rangle _0^J\mid _{J=0}=-ie\langle \overline{\psi }(x)\Bbb{\gamma }%
^i\psi (x)\rangle _0^J\mid _{J=0},  \tag{3.5}
\end{equation}
where $\rho _e$ and $j_e^i$ are the electric charge density and the electric
current density, respectively, in nuclear matter. We also may discuss the
current by means of Ref.\cite{huan}'s analogous method. Therefore, we obtain
the average relation of nuclear matter density and electric charge density
at the situation without external source as follows
\begin{equation}
\rho _g\equiv \langle \rho _B(x)\rangle _0^J\mid _{J=0}=\frac{-c}e\langle
\rho _e(x)\rangle _0^J\mid _{J=0},  \tag{3.6}
\end{equation}
where $\rho _{g\text{ }}$is the ground state density of the fermi doublet,
and $\rho _B(x)=\psi ^{+}(x)\psi (x)$ is the density operator of the proton
and neutron isospin doublet, Eq.(3.6)'s physical meaning is that the ground
state of nucleon density equates to the condensation of the electric charge
density divided by electronic charge and multiplied by $-c$, the
condensation is the distribution of the ground state density of charged
particles in nucleons.

We further get

\begin{equation}
\frac ie\langle j_e^i\rangle _0^J\mid _{J=0}=\frac ie\langle \partial _\nu
F^{i\nu }(x)\rangle _0^J\mid _{J=0}=\langle j^i\rangle _0^J\mid _{J=0}\equiv
j_0^i,  \tag{3.7}
\end{equation}
where $j^i=\overline{\psi }(x)\Bbb{\gamma }^i\psi (x)$ is a vector current
density of the nuclear matter.

On the other hand, because the interactions of $U_{EM}(1)$ and $SU_C(3)$
gauge fields generally affect the state of the matter, when the
corresponding external sources $J_{gauge}=(J_{A_\mu },J_{A_\mu ^\kappa
})\neq 0$ $(J_{A_\mu }$ and $J_{A_\mu ^\kappa }$ are external sources of the
interactions of $U_{EM}(1)$ and $SU_C(3)$ gauge fields, respectively, $%
\kappa \ $is $SU_C(3)$ color gauge group index ), we may generally assume a
general equivalent velocity $\mathbf{v}$ ( of the nuclear matter system with
$J_{gauge}\neq 0$) relative to the primordial (or called, ground state$%
^{\prime }$s) nuclear matter system with $J_{gauge}$ = 0, because the
equivalent relative velocity $\mathbf{v}$ is originated from the external
sources $J_{gauge}=(J_{A_\mu },J_{A_\mu ^\alpha })$ wth Lorentz
subscriptions. In fact, the actions of the external sources make the nuclear
matter system with $J_{gauge}\neq 0$ have the excited equivalently relative
velocity $\mathbf{v}$. Therefore, the velocity $\mathbf{v}$ is the function
of the external sources, i. e., $\mathbf{v=v(}J_{A_\mu },J_{A_\mu ^\kappa })=%
\mathbf{v(}J_{gauge}).$ Using a general Lorentz transformation we can obtain
the relations of the four dimensional general current of nuclear matter
system ( with $J_{gauge}\neq $0 ) relative to the nuclear matter system (
with $J_{gauge}$= 0 ) of the ground state as follows

\begin{equation}
\mathbf{j}^{\prime }=\mathbf{j}_0+\mathbf{v(}J_{gauge})\left[ (\frac 1{\sqrt{%
1-\frac{\mathbf{v}^2\mathbf{(}J_{gauge})}{c^2}}}-1)\frac{\mathbf{j}_0\mathbf{%
\cdot v(}J_{gauge})}{c^2}-\frac{\rho _g}{\sqrt{1-\frac{\mathbf{v}^2\mathbf{(}%
J_{gauge})}{c^2}}}\right] ,  \tag{3.8}
\end{equation}

\begin{equation}
\rho ^{\prime }=\frac{\rho _g-\frac{\mathbf{j}_0\text{ }\mathbf{\cdot }\text{
}\mathbf{v}\text{ }\mathbf{(}J_{gauge})}{c^2}}{\sqrt{1-\frac{\mathbf{v}^2%
\mathbf{(}J_{gauge})}{c^2}}}.  \tag{3.9}
\end{equation}

We, thus, can generally assume the velocity $\mathbf{v(}J_{gauge})$ linearly
depends on the external sources. Therefore, we can obtain a general
expression

\begin{equation}
\mathbf{v(}J_{gauge})=\mathbf{\alpha }_{A_\mu }J_{A_\mu }+\mathbf{\alpha }%
_{A_\mu ^\kappa }J_{A_\mu ^\kappa }  \tag{3.10}
\end{equation}
in which $\mathbf{\alpha }_{A_\mu }$ and $\mathbf{\alpha }_{A_\mu ^\kappa
\text{ }}$are the corresponding relative coupling constants of external
sources $J_{A_\mu }$ and $J_{A_\mu ^\kappa },$ respectively. Thus, Eqs.(3.8)
and (3.9) may be rewritten as two general expressions linearly depending on
the external sources as follows
\begin{equation}
\mathbf{j}^{\prime }=\mathbf{j}_0+(\mathbf{\alpha }_{A_\mu }J_{A_\mu }+%
\mathbf{\alpha }_{A_\mu ^\kappa }J_{A_\mu ^\kappa })\{(\frac 1{\sqrt{1-\frac{%
(\mathbf{\alpha }_{A_\mu }J_{A_\mu }+\mathbf{\alpha }_{A_\mu ^\kappa
}J_{A_\mu ^\kappa })^2}{c^2}}}  \tag{3.11}
\end{equation}

\[
-1)\frac{\mathbf{j}_0\mathbf{\cdot }(\mathbf{\alpha }_{A_\mu }J_{A_\mu }+%
\mathbf{\alpha }_{A_\mu ^\kappa }J_{A_\mu ^\kappa })}{c^2}-\frac{\rho _g}{%
\sqrt{1-\frac{(\mathbf{\alpha }_{A_\mu }J_{A_\mu }+\mathbf{\alpha }_{A_\mu
^\kappa }J_{A_\mu ^\kappa })^2}{c^2}}}\},
\]

\begin{equation}
\rho ^{\prime }=\frac{\rho _g-\frac{\mathbf{j}_0\text{ }\mathbf{\cdot }\text{
}(\mathbf{\alpha }_{A_\mu }J_{A_\mu }+\mathbf{\alpha }_{A_\mu ^\kappa
}J_{A_\mu ^\kappa })}{c^2}}{\sqrt{1-\frac{(\mathbf{\alpha }_{A_\mu }J_{A_\mu
}+\mathbf{\alpha }_{A_\mu ^\kappa }J_{A_\mu ^\kappa })^2}{c^2}}}.  \tag{3.12}
\end{equation}
and the consistent condition is

\begin{equation}
\left| \mathbf{\alpha }_{A_\mu }J_{A_\mu }+\mathbf{\alpha }_{A_\mu ^\kappa
}J_{A_\mu ^\kappa }\right| <c  \tag{3.13}
\end{equation}

In order to make the theory concrete, we consider a case when the external
source $J_{A_\mu ^\kappa }$ equates to zero but external $J_{A_\mu }$. Then
we gain the general case that there exists electromagnetic field, Eqs.(3.11)
and (3.12), thus, can be represented as

\begin{equation}
\mathbf{j}^{\prime }=\mathbf{j}_0+\mathbf{\alpha }_{A_\mu }J_{A_\mu
}\{(\frac 1{\sqrt{1-\frac{(\mathbf{\alpha }_{A_\mu }J_{A_\mu })^2}{c^2}}}-1)%
\frac{\mathbf{j}_0\mathbf{\cdot \alpha }_{A_\mu }J_{A_\mu }}{c^2}-\frac{\rho
_g}{\sqrt{1-\frac{(\mathbf{\alpha }_{A_\mu }J_{A_\mu })^2}{c^2}}}\} ,
\tag{3.14}
\end{equation}

\begin{equation}
\rho ^{\prime }=\frac{\rho _g-\frac{\mathbf{j}_0\text{ }\mathbf{\cdot }\text{
}\mathbf{\alpha }_{A_\mu }J_{A_\mu }}{c^2}}{\sqrt{1-\frac{(\mathbf{\alpha }%
_{A_\mu }J_{A_\mu })^2}{c^2}}},  \tag{3.15}
\end{equation}
and the corresponding consistent condition is $\left| \mathbf{\alpha }%
_{A_\mu }\right| <\frac c{J_{A_\mu }}$. When $\mathbf{\alpha }_{A_\mu }$ is
generally chosen as the motion direction $\mathbf{e}_x$, and $J_{A_{\mu
\text{ }}}$is taken as magnetic field B, we, thus, can have

\begin{equation}
j_x^{\prime }=\frac{j_{0x}-\rho _g\alpha B}{\sqrt{1-\frac{(\alpha B)^2}{c^2}}%
} ,  \tag{3.16}
\end{equation}

\begin{equation}
\rho ^{\prime }=\frac{\rho _g-\frac{\text{ }\alpha B}{c^2}j_{0x}}{\sqrt{1-%
\frac{(\alpha B)^2}{c^2}}},  \tag{3.17}
\end{equation}
where $\alpha $ is the small parameter determined by the nuclear physical
experiments under the external magnetic field B.

In order to test the theory, considering the case of $j_{0x}=0,$ in
Eq.(3.17) we have

\begin{equation}
\rho ^{\prime }=\frac{\rho _g}{\sqrt{1-\frac{(\alpha B)^2}{c^2}}}.
\tag{3.18}
\end{equation}
Because $\alpha $ is the coupling parameter, Eq(3.18) shows the relation of
density $\rho ^{\prime }$s coupling effect with external magnetic field,
which conforms to Ref.\cite{chak}$^{\prime }$s research about dense nuclear
matter in a strong magnetic field.

\section{Different Mass Spectrum about Different Dynamical Breaking and
Vacuum Breaking}

Because $\sigma _0$ and $\mathbf{\pi }_0$ may be made from the condensations
of fermion-antifermion, we can discuss the concrete expressions of different
mass spectrum about different dynamical breaking and different spontaneous
vacuum symmetry breaking as follows:

( i ) When considering the following dynamical breaking
\begin{equation}
\langle \overline{\psi }(x)\psi (x)\rangle _0^J\mid _{J=0}\neq 0,\text{ }%
\langle \overline{\psi }(x)\Bbb{\gamma }_5\mathbf{\tau }\psi (x)\rangle
_0^J\mid _{J=0}=0,  \tag{4.1}
\end{equation}
we has
\begin{equation}
\mathbf{\pi }_0=0,\text{ }\lambda \sigma _0(\nu ^2-\sigma _0^2)=g\langle
\overline{\psi }(x)\psi (x)\rangle _0^J\mid _{J=0}=-gtrS_F(0),  \tag{4.2}
\end{equation}
the corresponding spontaneous vacuum symmetry breaking is

\begin{equation}
\sigma (x)\longrightarrow \sigma (x)+\sigma _0,  \tag{4.3}
\end{equation}
the Lagrangian density, thus, is

\begin{align}
\mathcal{L}&=&-\frac 14F_{\mu \nu }F^{\mu \nu }-\overline{\psi
}(x)[\gamma
^\mu (\partial _\mu -ieA_\mu )+m_f]\psi (x)-g\overline{\psi }(x)[\sigma (x)+i%
\mathbf{\tau \cdot \pi }(x)\Bbb{\gamma }_5]\psi (x)]  \tag{4.4} \\
&&-\frac 12(\partial _\mu \sigma (x))^2-\frac 12m_\sigma ^2\sigma
^2(x)-\frac 12(\partial _\mu +ieA_\mu )\mathbf{\pi }^{+}(x)\cdot (\partial
_\mu -ieA_\mu )\mathbf{\pi }(x)-\frac 12m_\pi ^2\mathbf{\pi }^2(x)  \nonumber
\\
&&-\frac \lambda 4(\sigma ^2(x)+\text{ }\mathbf{\pi }(x))^2-\lambda \sigma
_0\sigma (x)(\sigma ^2(x)+\mathbf{\pi }^2(x))-gtrS_F(0)\sigma (x).  \nonumber
\end{align}

One obtains that the fermion doublet masses are

\begin{equation}
m_f=g\sigma _0.  \tag{4.5}
\end{equation}

Masses of $\sigma (x)$ and $\mathbf{\pi }(x),$ respectively, are

\begin{equation}
m_\sigma ^2=\lambda (3\sigma _0^2-\nu ^2)=2\lambda \sigma
_0^2+gS_F(0)/\sigma _0,  \tag{4.6}
\end{equation}

\begin{equation}
m_\pi ^2=\lambda (\sigma _0^2-\nu ^2)=gtrS_F(0)/\sigma _0.  \tag{4.7}
\end{equation}
Thus, when there is no dynamical breaking, we obtain $\sigma _0^2=\nu ^{2%
\text{ }},$ which just shows $\nu ^2$'s physical meaning, i.e., $\sigma _{0%
\text{ }}$is, in this case, just the spontaneous vacuum breaking parameter,
and $m_\pi ^2=0$. Even so, $\sigma $ particles and fermions acquire masses,
namely $m_\sigma ^2=2\nu ^2,$ $m_f=g\left| \nu \right| .$ Therefore, the
masses of $\sigma $ particle and fermion doublet naturally come from only
the vacuum breaking structure. In general case, when there exist both
dynamical breaking and spontaneous vacuum breaking, not only $\mathbf{\pi }$
meson and fermions gain masses, but also $\sigma $ and $\mathbf{\pi }$
masses are not equal. More generally, we may take $\sigma _0^{\prime
}=\langle \overline{\psi }(x)\psi (x)\rangle _{0\text{ }}^J$in which $\sigma
_{0\text{ }}^{\prime }$is the running spontaneous vacuum breaking value. It
means that $\sigma _{0\text{ }}^{\prime }$is the exciting state, which make
fermion doublet, $\sigma $ particle and $\mathbf{\pi }$ gain effective
masses relative to different external sources.

( ii ) when $\sigma _0=0$ , $\pi _0=0$ , analogous to the research about
Eqs.(4.6) and (4.7), we get $\sigma (x)$ and $\mathbf{\pi }(x)$ meson having
the same mass\cite{cla}

\begin{equation}
m_\sigma ^2=m_\pi ^2=-\lambda \nu ^2.  \tag{4.8}
\end{equation}

Further using Eq.(4.5) in the cases of $\sigma _0=0$ and $\pi _0=0$, we
obtain the fermion doublet keeping no mass.

( iii ) General dynamical breaking

We now consider a general dynamical breaking. From Eqs.(3.2) and (3.3) we
see that

\begin{equation}
\sigma _0=Kg\langle \overline{\psi }(x)\psi (x)\rangle _0^J\mid _{J=0}\neq 0,%
\text{ }\mathbf{\pi }_0=iKg\langle \overline{\psi }(x)\Bbb{\gamma }_5\mathbf{%
\tau }\psi (x)\rangle _0^J\mid _{J=0}\neq 0.  \tag{4.9}
\end{equation}

Then the corresponding spontaneous vacuum symmetry breaking are

\begin{equation}
\sigma (x)\longrightarrow \sigma (x)+\sigma _0,\text{ }\mathbf{\pi }%
(x)\longrightarrow \mathbf{\pi }(x)+\varepsilon \mathbf{\pi }_0\text{, }%
0\leq \varepsilon \leq 1,  \tag{4.10}
\end{equation}
where $\varepsilon $ is a running breaking coupling parameter determined by
different physical experiments.

Because electromagnetic interaction is very weaker than strong interaction,
electromagnetic interaction may be neglected. The corresponding Lagrangian,
is

\[
\mathcal{L}=-\overline{\psi }(x)[\gamma ^\mu \partial _\mu +m_f]\psi (x)-g%
\overline{\psi }(x)[\sigma (x)+i\mathbf{\tau \cdot \pi }(x)\Bbb{\gamma }%
_5]\psi (x)-\frac 12(\partial _\mu \sigma (x))^2-\frac{m_\sigma ^2}2\sigma
^2(x)-\frac 12(\partial _\mu \mathbf{\pi }(x))^2
\]

\[
-\frac \lambda 2[(\sigma _0^2+\varepsilon ^2\mathbf{\pi }_0^2-\nu ^2)\mathbf{%
\pi }^2+2(\varepsilon \mathbf{\pi }_{0\cdot }\mathbf{\pi })^2]-\frac \lambda
4(\sigma ^2(x)+\text{ }\mathbf{\pi }^2(x))^2-\lambda (\sigma _0\sigma
(x)+\varepsilon \mathbf{\pi }_{0\cdot }\mathbf{\pi }(x))(\sigma ^2(x)+%
\mathbf{\pi }^2(x))
\]
\begin{equation}
-2\lambda \sigma _0(\varepsilon \mathbf{\pi }_0\cdot \mathbf{\pi }(x))\sigma
(x)-\frac \lambda 2(\sigma _0^2+\varepsilon ^2\mathbf{\pi }_0^2-\nu
^2)(\sigma _0\sigma (x)+\varepsilon \mathbf{\pi }_{0\cdot }\mathbf{\pi }%
(x))-\frac \lambda 4(\sigma _0^2+\varepsilon ^2\mathbf{\pi }_0^2-\nu ^2)^2,
\tag{4.11}
\end{equation}
where masses of the fermions and $\sigma $ particle , respectively, are

\begin{equation}
m_N=g(\sigma _0+i\varepsilon \mathbf{\tau \cdot \pi }_0\Bbb{\gamma }_5) ,
\tag{4.12}
\end{equation}

\begin{equation}
m_\sigma ^2=\lambda (3\sigma _0^2+\varepsilon ^2\mathbf{\pi }_0^2-\nu ^2) .
\tag{4.13}
\end{equation}

Because of
\begin{equation}
\left( {\pi _0 \cdot \pi } \right)^2=\pi _0^2 \pi
^2+\sum_{\stackrel{i,j=1}{i\not=j}}^3 {\left( {\pi _{i0} \pi _{j0}
\pi _i \pi _j -\pi _{i0}^2 \pi _j^2 } \right)}\tag{4.14}
\end{equation}

Under the condition of $ \sum_{\stackrel{i,j=1}{i\not=j}}^3\pi
_{i0}\pi _{j0}\pi _i\pi _j= \sum_{\stackrel{i,j=1}{i\not=j}}^3\pi
_{i0}^2\pi _j^2 \ $, we obtain meson mass expression

\begin{equation}
m_\pi ^2=\lambda (\sigma _0^2+3\varepsilon ^2\mathbf{\pi }_0^2-\nu ^2).
\tag{4.15}
\end{equation}

When $\mathbf{\pi }_0=0$ or $\varepsilon =0$, the results (iii) are
simplified into the results (i) above.

When there is pseudoscalar condensation $\langle \overline{\psi }(x)\mathbf{%
\tau }\Bbb{\gamma }_5\psi (x)\rangle _0^J\mid _{J=0},$ because the scalar
condensation is stronger than the pseudoscalar condensation, the $\sigma
_{0_{\text{ }}\text{ }}$is not equal to zero under existing pseudoscalar
condensation.

From the above discussion, we may see what no needing Higgs particle, we
naturally gain both fermion$^{\prime }$s masses and boson$^{\prime }$s ($%
\sigma $ and $\mathbf{\pi }$ ) masses. The mechanisms of gaining masses are
more direct and useful for constructing the weak-electromagnetic standard
model without Higgs fields. For making fermions and bosons in the other
models acquire masses, it may make the too many adjusting parameters of
fitting with the physical experiments in the usual unified models decrease.
We, further, generally deduce that the masses of nucleons, $\sigma $ and $%
\mathbf{\pi }$ have the effects coming from interactions with external
source. It can be seen that $\sigma $ and $\pi ^{0\text{ }}$may be made from
the different condensations of fermion and antifermion. This lead to that $%
\sigma $ and $\pi ^0$ without electric charge have electromagnetic
interaction effects coming from their inner construction. Using the
all general research of this paper$\mathbf{,}$we can very more study
the interactions between different fundamental particles in general
situation, all these will be written in the other papers.

\section{Summary and Conclusion}

We show up a general $SU(2)_L\times SU(2)_R$$\times U(1)_{EM}$ $\sigma $-
model with external sources, dynamical breaking and spontaneous vacuum
symmetry breaking. We present the general basic formulations of the model.
This paper founds the different condensations about fermions and
antifermions in which the concrete scalar and pseudoscalar condensed
expressions of $\sigma _{0\text{ }}$and $\mathbf{\pi }_{0\text{ }}$bosons
are shown up. We have shown that $\sigma $ and $\pi ^{0\text{ }}$may be made
from the different condensations of fermion and antifermion. We have
discovered that $\sigma $ and $\pi ^{0\text{ }}$without electric charge have
electromagnetic interaction effects coming from their inner construction,
which is similar to neutron. Using a general Lorentz transformation and four
dimensional condensed currents of the nuclear matter of the ground state
with J = 0 we deduced the four dimensional general relations of different
currents of the nuclear matter system with J $\neq 0$ relative to the ground
state$^{\prime }$s nuclear matter system with J = 0, and give the relation
of density $\rho ^{\prime }$s coupling effect with external magnetic field.
This conforms to Ref.\cite{chak}'s research about dense nuclear matter in a
strong magnetic field. We also get the concrete expressions of different
mass spectrum about different dynamical breaking and spontaneous vacuum
breaking. This paper has given running spontaneous vacuum breaking value $%
\sigma _{0\text{ }}^{\prime }$in terns of the technique of external sources,
has obtained spontaneous vacuum symmetry breaking based on the $\sigma
_0^{\prime }$, which make nuclear fermion doublet, $\sigma $ and $\mathbf{%
\pi }$ particles gain effective masses relative to external sources. We have
found out the mechanisms of mass production of fermion doublet and bosons ($%
\sigma $ and $\mathbf{\pi }$ ). The mechanism is useful for constructing the
unified weak-electromagnetic model without fundamental scalar fields. The
effect of external sources and nonvanishing values of the scalar and
pseudoscalar condensations are given in this theory, we generally deduce
that the masses of nucleons, $\sigma $ and $\mathbf{\pi }$ partly come from
the interactions with different external sources.

Acknowledgment: The authors are grateful to Prof. Z. P. Li for useful
discussion.

This work was partially supported by the CAS knowledge Innovation Project
(No.KJCX2-sw-No2) and Ministry of Science and Technology of People's
Republic of China (No.2002ccb 00200).

\end{document}